# Identification of image source using serial-number-based watermarking under Compressive Sensing conditions

Andjela Draganić*, Milan Marić**, Irena Orović* and Srdjan Stanković*
* University of Montenegro, Faculty of Electrical Engineering, Podgorica, Montenegro
** S&T Crna Gora d.o.o, Podgorica, Montenegro

**Abstract** – Although the protection of ownership and the prevention of unauthorized manipulation of digital images becomes an important concern, there is also a big issue of image source origin authentication. This paper proposes a procedure for the identification of the image source and content by using the Public Key Cryptography Signature (PKCS). The procedure is based on the PKCS watermarking of the images captured with numerous automatic observing cameras in the Trap View cloud system. Watermark is created based on 32-bit PKCS serial number and embedded into the captured image. Watermark detection on the receiver side extracts the serial number and indicates the camera which captured the image by comparing the original and the extracted serial numbers. The watermarking procedure is designed to provide robustness to image optimization based on the Compressive Sensing approach. Also, the procedure is tested under various attacks and shows successful identification of ownership.

## I. Introduction

Development of new ICT technologies improved the ease of creation and access to digital information. However, ease access to digital information increases the doubt about who is the creator or owner of the digital content, as well as its authenticity. Namely, by using various digital tools, digital media copies can be distributed with changed information about source, author or owner. The public key digital signature (PKCS), embedded into the digital media, can provide services for strong origin authentication and reliable content integrity of digital image. PKCS technology is reliable and strong in terms of protection on hacker attacks, but it is fragile for use in environments where digital images are subject to modification or exposed to communication noise during transport to receiver. Therefore, this paper proposes the watermarking procedure that aims in preserving the embedded part of the PKCS, even if the digital image is exposed to the various attacks.

Watermarking techniques provides indication of ownership, and/or indication of the identity of a licensed user, by embedding the information in the digital object [1]-[12]. This information may be visible or hidden, but the security properties of this technology are limited due to possible malicious attacks on the original image. Image watermarking techniques usually embed the security information throughout the digital pixels in a manner that does not impact its normal use [1],[11]-[13]. There is always the requirement that the watermark should be capable of surviving routine transformations to the image, such as blurring, cutting or compression. Adding digital signature in cover work of the watermarking techniques could highly improve safeguarding against malicious source origin change and unauthorized image modification.

In the proposed procedure, part of the PKCS – a serial number (SN), is used for embedding. The SN, that is available in the hexadecimal form, is transformed into the 32-bit binary form. Based on the binary form, logo is created and embedded into the image as a watermark. Watermarked image is transmitted through the network and may be exposed to the attacks. At the receiver side, the watermark detection extracts 32-bit binary sequence. This sequence is then compared with the original sequence, i.e. with the SN corresponding to the camera device that captured the observed image. This original SN is available through the Certification Authority system.

The procedure of watermarking and camera identification is tested on the images from the TrapView pest monitoring system [14],[15]. The TrapView system is consisted of traps distributed through the fields/orchards with a purpose of monitoring pests captured in those traps. The images of caught insects are sent regularly to the TrapView cloud that later provides pest recognition and pest occurrence statistics. When the number of insect specimens becomes too large, trap has to be renewed. The SN-based watermarking procedure helps to locate the trap that has to be renewed, by revealing the SN of the camera device.

Having in mind that the images has to be regularly uploaded to the cloud, and that they are of high resolution,

the image size optimization has to be done prior it is sent over mobile network. The optimization is done by using the Compressive Sensing (CS) approach [16]-[23]. By randomly selecting only small number of image samples from its frequency domain, the image can be successfully reconstructed at the receiver size by using an optimization algorithms. The SN has to be preserved after the random samples selection and optimization. In other words, the proposed watermarking procedure should be robust to the CS attack.

The paper is organized as follows. The TrapView system and image optimization by using the CS is described in the Section 1. Theoretical background on the image watermarking is provided in the Section 2. Section 3 describes the proposed watermarking procedure, while in Section 4 the experiments are given. Conclusion is in the Section 5.

## II. THEORETICAL BACKGROUND

TrapView pest monitoring system is a platform that provides information about occupancy of traps distributed through the fields/orchards. The system captures and uploads the images to the TrapView cloud, at daily basis. An automated pest monitoring is useful in cases when there is need for monitoring large areas. Each TrapView camera has its own PKCS, and part of the PKCS is the SN. System is illustrated in Figure 1.

In order to be able to detect which camera captured the observed image, the SN is embedded into the image. The SN embedding can be added to the standard TrapView system after image capturing, as it is shown in the Figure 1. As the TrapView captured image is around 1MB large, our goal is to reduce its size without loss of quality and by preserving the SN embedded in each captured image.

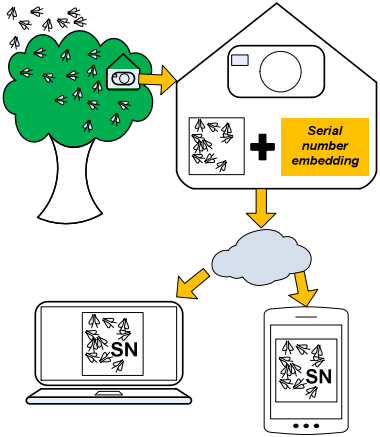

Figure 1. TrapView system with SN embedding

Let us firstly describe the size optimization approach. The optimization is done by applying the CS method. The CS aims at recovering data from the small set of available samples. Signal can be intentionally under-sampled, or samples corrupted by noise or some other environmental factors can be considered as missing. Random selection of the signal coefficients, in the domain where signal has dense representation, assures successful reconstruction from the small number of acquired samples. If an $N$-dimensional signal $\mathbf{x}$ has a sparse representation in the certain transform domain $\mathbf{\Psi}$:

$$\mathbf{x} = \sum_{i=1}^{N} \mathbf{X}_i \psi_i = \mathbf{\Psi}\mathbf{X}, \quad (1)$$

where $\mathbf{X}_i$ is a transform domain coefficient and $\psi_i$ is a basis vector, then the vector of acquired samples $\mathbf{y}$ is defined as:

$$\mathbf{y} = \mathbf{\Phi}\mathbf{x} = \mathbf{\Phi}\mathbf{\Psi}\mathbf{X} = \mathbf{A}\mathbf{X}. \quad (2)$$

The matrix $\mathbf{A}$ denotes the measurement (CS) matrix, while the matrix $\mathbf{\Phi}$ models random selection of the coefficients. The signal is reconstructed by solving the set of linear equations (2), using an optimization algorithms, i.e. finding the sparsest solution of the system. In the case of 2D data, commonly applied method for the optimization problem solving is the TV optimization, based on minimization of the image gradient.

For solving the system (2), the minimization over $\mathbf{X}$ of the regularization function $J(\mathbf{X})$ is performed:

$$J(\mathbf{X}) = \frac{\mu}{2} \|\mathbf{y} - \mathbf{A}\mathbf{X}\|^2 + \lambda R(\mathbf{X}), \quad (3)$$

where $\lambda \in (0, \infty)$, $R(\mathbf{X})$ is the TV of the signal $\mathbf{X}$:

$$R(\mathbf{X}) = \|d\mathbf{X}\|_{\ell_1},$$
$$d_{i,j}\mathbf{X} = \begin{bmatrix} \mathbf{X}(i+1, j) - \mathbf{X}(i, j) \\ \mathbf{X}(i, j+1) - \mathbf{X}(i, j) \end{bmatrix}, \quad (4)$$

and $d$ is a gradient operator.

The TV optimization is applied on the TrapView images. The TrapView camera, placed in the field, takes pictures of insects captured in the trap and sends them to the cloud. Before sending, the image is transformed in the discrete cosine transform (DCT) domain and under-sampled. Therefore, the vector of measurements $\mathbf{y}$ is consisted of the DCT coefficients. The optimization problem is defined as:

$$\min_{\mathbf{X}} \mathrm{TV}(\mathbf{X}) \text{ subject to } \mathbf{y} = \mathbf{A}\mathbf{X}, \quad (5)$$

Or, in the discrete form:

$$\mathrm{TV}(\mathbf{X}) = \sum_{i,j} \sqrt{(\mathbf{X}_{i+1,j} - \mathbf{X}_{i,j})^2 + (\mathbf{X}_{i,j+1} - \mathbf{X}_{i,j})^2}. \quad (6)$$

## III. THE PROCEDURE FOR TRAPVIEW IMAGES WATERMARKING AND CS OPTIMIZATION

### A. TrapView image watermarking

The watermarking procedure is based on embedding of the SN, part of the digital certificate, that corresponds to certain camera device. Digital certificates are issued from the Certification Authority (CA), and each issued certificate has its own and unique SN, validity period and a subject to whom is issued. The SN provides the name or source origin of the image, stored in register of digital certificates issued by CA, and consequently identifies the camera which is the source of the captured image.

The SN consists of 32 bits grouped in 8 hex numbers, thus forming unique digital identifier for a subject. In the proposed procedure, the group of 8 hex numbers is firstly modified from its original form and binary sequence is created, as it is shown in the Figure 2.

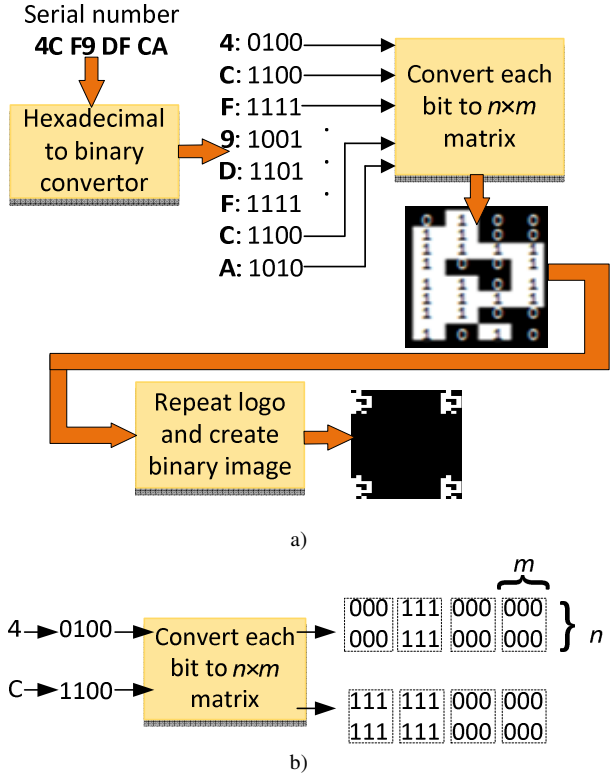

Figure 2: a) Logo creation for the SN example "4CF9DFCA"; b) simplified illustration of the bit-to-matrix conversion

Each symbol is represented with 4 bits, which results in 32-bit sequence. After hex-to-binary conversion, each bit from the sequence is represented in the $n \times m$ matrix form. Therefore, one hex symbol forms an $n \times 4m$ matrix (see Figure 2). After all of the 8 symbols are converted into the matrix form, the binary logo image is created.

In order to increase robustness to different watermark attacks, the logo image is repeated and new binary image with repeated logos is created. This binary image is of the same size as the original image. The starting logo is chosen to be embedded 4 times, in the image corners, as it is shown in Figure 2.

The next step is logo embedding. The original logo is spread into the several bit planes, and only part of the logo is embedded into each plane. If we assume that the image coefficient is represented with $B$-bits, then $B$ bit planes are available. We chose $L=4$ bit planes for embedding and four $n \times 4m$ matrices in the image corners are used. Middle bit planes are considered [9], in order to provide robustness and, at the same time, to avoid the influence of the watermark to the image quality. The logo is divided into the $L$ parts (layers), by using a unique security key, i.e. a unique random matrix $\Re$ [9]. The matrix is separated on $L$ layers, and each layer contains values from the certain interval - there are $L$ non-overlapping intervals. Therefore, the random matrix $\Re$ and the threshold values, together with selection of the bit planes, form a security key. The complete logo is obtained after summation of the $L$ layers (Figure 3).

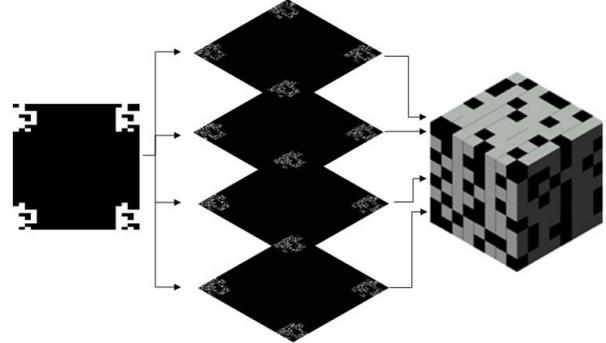

Figure 3: Logo separation and embedding into the bit planes

If $W_k$ denotes the $k$-th layer of the logo $W$, $B$ number of bit planes, $M$ and $N$ are image sizes and $\omega_k$ are the threshold values for the layers, then following logo creation procedure can be defined:

***Procedure for logo creation:***

$k \in \{1,...,B\}$
- for $i=1:M$
-   for $j=1:N$
-     if $\omega_{k-1} < \Re(i, j) < \omega_{k-1}$ then
-       $W_k(i,j)=W(i,j)$
-     else
-       $W_k(i,j)=0$
-     end if
-   end for
- end for

The threshold values are chosen by using an equidistant rule, which is one way of secrecy providing. In our case, the elements of the random matrix $\Re$ take values from the interval $(0,1)$ and the threshold values for the corresponding layers are $(0, 1/4)$, $(1/4, 1/2)$, $(1/2, 3/4)$ and $(3/4, 1)$. If we denote the observed image with **x**, the embedding can be described as:

***Procedure for logo embedding:***

$k \in \{1,...,B\}$
- for $i=1:M$
-   for $j=1:N$
-     if $W_k(i,j)=1$ then
-       $x(i,j)=W_k(i,j)$
-     else
-       $x(i,j)=x(i,j)$
-     end if
-   end for
- end for

The watermark extraction is done on the receiving side. Depending on the attack to which image is exposed during transmission, the logo will be degraded at certain degree. The SN is extracted from the received logo. Since the logo is consisted of 4 small logos at the image corners, the 4 SN will be extracted. If we find at least one SN corresponding to the embedded original SN, and if we have successful comparison between original and received SN, we are proving the source origin. The details for the original SN are requested from the Montenegro Public Certification Authority – Post CG CA.

The procedure for logo extraction starts by observing the bit planes where watermark is added, and can be described with the following procedure:

***Procedure for logo extraction:***

$k \in \{1,...,B\}$
- for $i=1:M$
-     for $j=1:N$
-         if $W(i,j)=1$ and $\omega_{k-1} < \Re(i,j) < \omega_k$ then
-             $W_k^{ext}(i,j) = W_k(i,j)$
-         else
-             $W_k^{ext}(i,j) = 0$
-         end if
-     end for
- end for

Logo extraction cannot be done unless the random matrix $\Re$, together with the thresholds and the bit plane order numbers, are known.

*B. CS as a watermark attack*

A special attention is devoted to the CS attack of the watermarked image. After the SN embedding, the CS is applied to the captured image, with an aim to decrease the amount of transmitted information. Certain percent of the pixels from the watermarked image is selected. Selection is done at a random manner. After that, the selected pixels are transmitted and the image is reconstructed at the receiver side. High image quality after the CS reconstruction is important for:
- image post-processing;
- SN extraction.

Post-processing of the images means counting the number of captured insect specimens, based on counting algorithms. Therefore, it is of great interest to save the quality of the reconstructed image as better as it is possible, in order to minimize the possibility of error occurrence during counting. As it was mentioned previously, an accurate SN extraction is important for the camera device location.

IV. EXPERIMENTAL RESULTS

Let us consider the SN-based watermarking of the TrapView images. The overview of digital certificate for certain user/device is shown in Figure 4.

The observed SN is in the hexadecimal form: 4C F9 DF CA. The first step of the proposed procedure is hexadecimal to binary conversion. Therefore, after the conversion, the following sequence is obtained: 0100 1100 1111 1001 1101 1111 1100 1010. The next step assumes forming matrix from each bit in the sequence. This part of the procedure results in binary logo formation, as it is shown in Figure 2a.

The logo is repeated 4 times (see Figure 2b), prior it is embedded into the 4 bit planes (the coefficients of the image are represented with 8 bits). In order to determine the positions of sub-images pixels used in watermarking procedure, the random matrix $\Re$ is exploited and it has to be known at the receiver side.

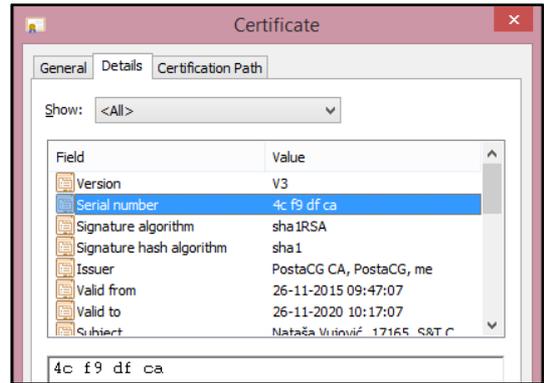

Figure 4: Overview of digital certificate with serial number details

The original and the watermarked images are shown in Figure 5. As it can be seen, the watermark does not degrade the image quality. In order to decrease the number of transmitted samples per image, as well as to increase transmission and upload speed, only 21% of randomly selected samples per image are chosen.

TV-based optimization is done at the receiver side and the image is reconstructed. After the optimization, the logo and SN are extracted. The CS reconstructed image is shown in Figure 6a, while the extracted logo is shown in Figure 6b. The obtained peak signal to noise ratio (PSNR) is 30.0082 dB, which numerically proves satisfactory reconstruction quality. Therefore, we can conclude that the CS approach will not affect the process of specimens counting.

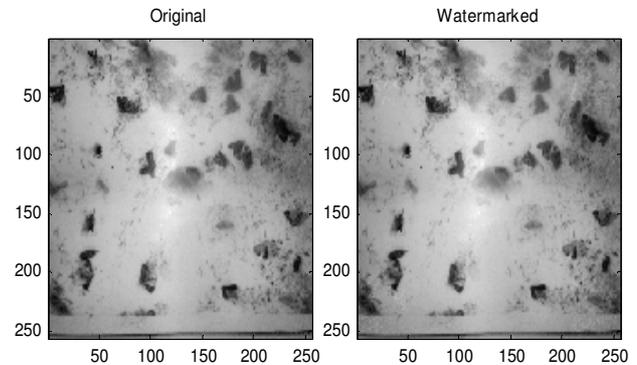

Figure 5: Original and watermarked image

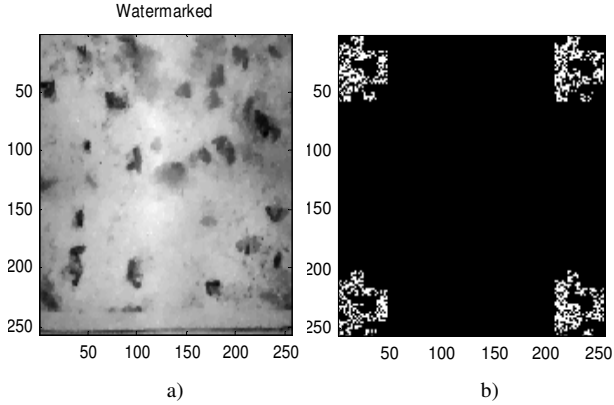

Figure 6: a) Image reconstructed using 21% of the total number of image samples; b) the extracted logo from the image shown in a)

*Table 1: Logo and SN extraction results for various attacks*

| Attack | Gaussian noise | Impulse noise |
|---|---|---|
| Extracted logo | | |
| Extracted SN | All 4 SN extracted | All 4 SN extracted |
| | PSNR=10.1167 Db | 40% of the image samples corrupted |
| Attack | JPEG compression, quality = 10 | CS with 21% available samples |
| Extracted logo | | |
| Extracted SN | All 4 SN extracted | All 4 SN extracted |
| Attack | Image brightening (80%) | Image darkening |
| Extracted logo | | |
| Extracted SN | All 4 SN extracted | All 4 SN extracted |
| Attack | Median filtering | Image blurring |
| Extracted logo | | |
| Extracted SN | All 4 SN extracted | All 4 SN extracted |

The SN is detected by using the extracted logo. Based on counting the "1" and the "0" in each $n \times 4m$ matrix and taking the value that correspond to the greater number of occurrences, the decision on serial number's bit is made and the extracted SN appears in binary form.

The SN extraction is done from all 4 logos, because if only one bit from 32 sequence is modified, the identification fails. Therefore, we check the extracted SN 4 times, and if match with the original SN is obtained just once, we can say that the camera device identification is successful. In this case, all 4 extracted logos provide the exact SN. Beside the CS, the robustness of the watermarking procedure is tested under other commonly appeared attacks in real applications. The success of the SN extraction is observed, and the results are given in the Table 1.

## V. CONCLUSION

The procedure for image watermarking, based on the Public Key Cryptography Signature and serial number embedding, is proposed in the paper. The images captured with an automatic observing cameras in the Trap View pest monitoring system, are used. Each camera has its own serial number, based on which the device identification is done. The images are watermarked in order to ease the camera identification that captured the observed image. Based on 32-bit PKCS serial number, the binary logo is created and embedded into the captured image. Detection of the watermark is done at the receiver side, and indicates the camera which captured the image by comparing the extracted serial number with an original one. The watermarking procedure is defined in a way that provides robustness to CS-based image optimization. The 80% of the image samples can be avoided and image can still be reconstructed, preserving the embedded serial number. The robustness of the procedure is also tested under other common watermarking attacks and shows successful serial number extraction in all considered cases.

## ACKNOWLEDGMENT

This work is supported by the Montenegrin Ministry of Science, project grant funded by the World Bank loan: CS-ICT "New ICT Compressive sensing based trends applied to: multimedia, biomedicine and communications". The authors are thankful to Mr. Matej Štefančič, the director of EFOS, for providing us test images.